\documentclass[doublecol]{epl2}
\usepackage[utf8x]{inputenc}
\usepackage{graphicx,amsmath,color,url}

\newcommand{\vek}[1]{\boldsymbol{#1}}

\title{The essential role of time in network-based recommendation}
\author{Alexandre Vidmer and Matúš Medo\footnote{matus.medo@unifr.ch}}
\shortauthor{Alexandre Vidmer \and Matúš Medo}
\institute{Department of Physics, University of Fribourg, Chemin du Musée 3, CH-1700 Fribourg, Switzerland}
\abstract{Random walks on bipartite networks have been used extensively to design personalized recommendation methods. While aging has been identified as a key component in the growth of information networks, most research has focused on the networks' structural properties and neglected the often available time information. Time has been largely ignored both by the investigated recommendation methods as well as by the methodology used to evaluate them. We show that this time-unaware approach overestimates the methods' recommendation performance. Motivated by microscopic rules of network growth, we propose a time-aware modification of an existing recommendation method and show that by combining the temporal and structural aspects, it outperforms the existing methods. The performance improvements are particularly striking in systems with fast aging.}


\begin{document}

\maketitle

\section{Introduction}
Increasing data availability and computational capacity~\cite{lazer2009life}, interconnections between previously separate data domains~\cite{dong2012link,bao2015recommendations}, and the immediate commercial importance of recommendation~\cite{schafer1999recommender,bell2007lessons,pathak2010empirical} all contribute to the unceasing interest in the study of recommender systems~\cite{bobadilla2013recommender}. The goal of recommendation is to use data on past user preferences to obtain personalized ``recommendation'' of new items (shopping items, YouTube videos, or any other content) that an individual user might appreciate. From the physics perspective, it has been interesting to realize that well-known physics processes, such as random walks and heat diffusion, on network representations~\cite{newman2010networks} of the underlying data give rise to efficient recommendation methods~\cite{zhang2007recommendation,zhou2007bipartite,zhou2010solving,yu2016network}.

Despite physics being a science that aims at understanding the evolution of systems, the research of network-based recommendation by physicists has entirely neglected the dimension of time which turns out to be of high importance for traditional recommendation approaches~\cite{ding2005time,koren2010collaborative,campos2014time,daneshmand2014time}. While this is understandable from the historical perspective---early datasets often lacked the time information---the situation is very different now. The role of time in the evolution of information networks (that serve as input data for recommendation) has been demonstrated~\cite{crane2008robust,szabo2010predicting,ren2016characterizing}, modeled~\cite{medo2011temporal,medo2014statistical,hou2014memory}, and turned into numerous useful applications~\cite{wang2013quantifying,zhou2015temporal,mariani2016quantifying}.

The ignorance of time in the research of network-based recommendation manifests itself in the evaluation of recommendation methods. This evaluation is normally done by hiding part of the input data---this part is commonly referred to as the \emph{probe} set---and using the rest of the data---which is referred to as the \emph{training} set---as input for a recommendation method. The obtained results are finally evaluated on the basis of how well they reproduce the hidden data~\cite{herlocker2004evaluating,shani2011evaluating}. Crucially, the probe is typically chosen at random~\cite{lu2012recommender,yu2016network}; the training set therefore includes both past and future data entries.

We show for the first time that when instead the latest data are hidden and thus the task is to reproduce strictly future data entries, the performance of network-based recommendation methods becomes dramatically worse than the performance observed on a random probe. Sophisticated network-based recommendation methods then turn out to be outperformed by trivial approaches such as the recent item popularity increase which has been shown to be a good predictor of the future popularity increase~\cite{zeng2013trend}. We propose to combine a simple network-based recommendation method with dynamical features of the network growth. The resulting hybrid method is shown to consistently and significantly outperform the existing methods on various e-commerce datasets.

\begin{figure}[t!]
\centering
\includegraphics[scale=0.63]{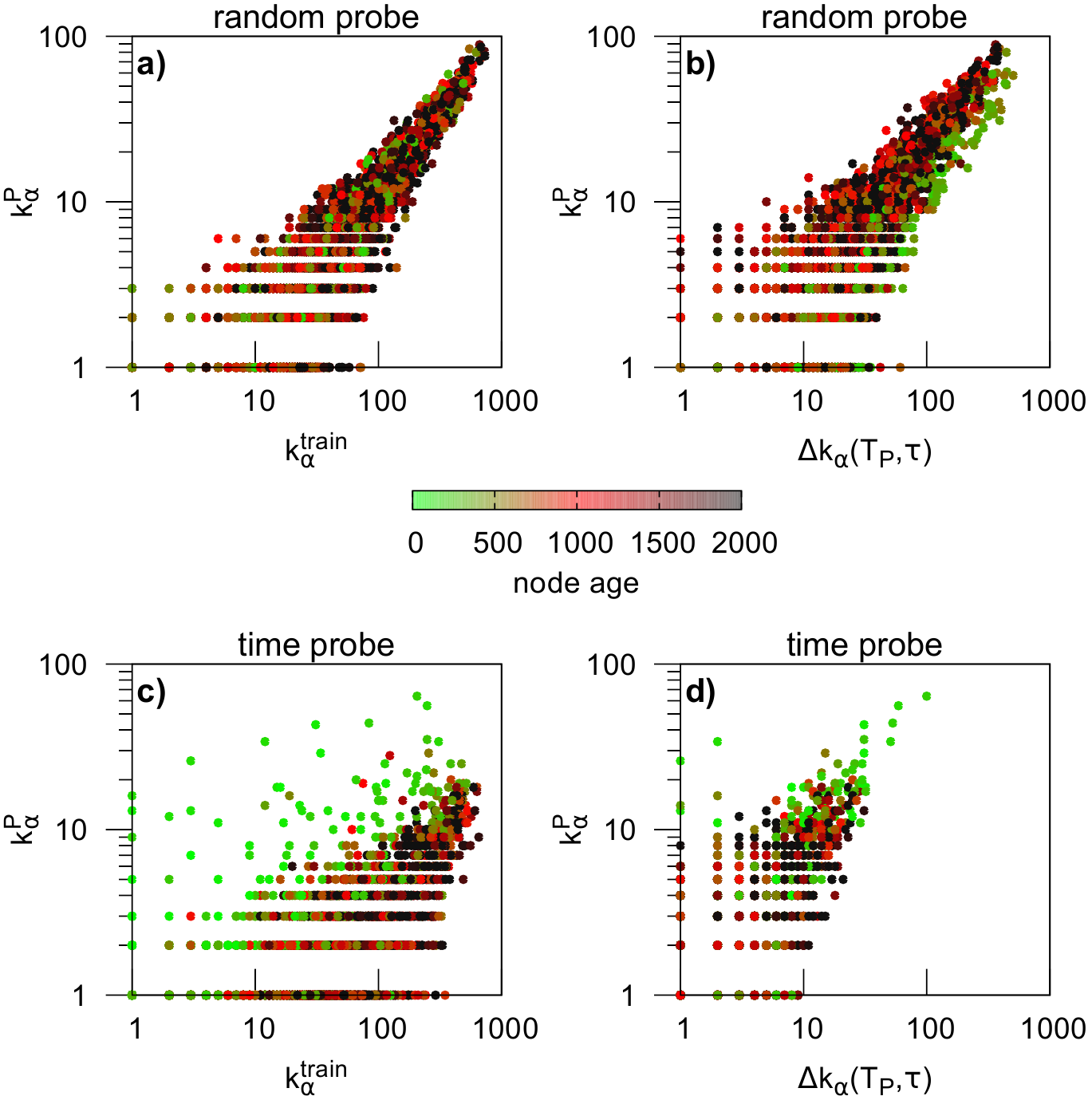}
\caption{Comparison of the random and time probe in the Netflix data. Correlation between item degree in the training data $k_{\alpha}^{\text{train}}$ and the number of item links in the probe $k_{\alpha}^P$ is higher in the random probe (panel a, Pearson correlation $0.97$) than in the time probe (panel c, Pearson correlation $0.70$). In the time probe, the dark-shaded nodes of small age achieve higher degree increase than green-shaded old nodes (panel c). Furthermore, recent degree increase $\Delta k_{\alpha}(T_P, \tau)$ correlates well with the number of item links in the time probe $k_{\alpha}^P$ (panel d, Pearson correlation $0.84$). Expectedly, this correlation is smaller in the random probe (panel b, Pearson correlation $0.66$). The color of each node-representing symbol corresponds to the node age. We use here $\Delta_P=\tau=20\,\text{days}$. Results for the two other datasets are similar. The Digg data features faster aging and thus the difference between the random and time probe is greater: the correlation between the training and probe degree is 0.99 and 0.02, respectively.}
\label{fig:dk_vs_k}
\end{figure}

\section{The problem of a random probe}
\label{sec:problem}
We begin by describing in detail how a probe chosen at random favors network-based recommendation methods. Input data for network-based recommendation consists of a set of links between users and items; links connect users with the items that they have collected, or otherwise associated themselves with (\emph{e.g.}, a user is connected with the products they have reviewed on Amazon.com). These data are naturally represented as a bipartite user-item network with $U$ user nodes and $I$ item nodes. Elements of the network's evolving adjacency matrix, $a_{i\alpha}(t)$, are one when user $i$ is connected with item $\alpha$ at time $t$ and zero otherwise. Degree $k_i(t)$ of user node $i$ corresponds to the number of items collected by user $i$ until time $t$, and degree $k_{\alpha}(t)$ of item node $\alpha$ corresponds to the number of users who have collected item $\alpha$ until time $t$. Both the user and item degree distribution are typically broad~\cite{clauset2009power,zhou2011emergence}.

Upon random probe division, links from the input data are chosen at random and moved to the probe. The probability that a randomly chosen link is attached to item $\alpha$ is proportional to its degree $k_{\alpha}(t)$. At the same time, most network-based recommendation methods are known to tend to recommend popular items~\cite{zhou2010solving,liu2011information,qiu2011item} which means that choosing the probe at random is inherently favorable to these methods because thus-created probes share the bias of recommendation methods themselves.

While the tendency of the random probe towards popular items may seem worrisome, one might think that due to the omnipresent preferential attachment mechanism~\cite{albert2002statistical} (Section~8), degree of an item is a good proxy for the item's future degree increase. If that would be the case, the random probe would be effectively equivalent to the probe consisting of the most recent links. However, aging has been shown to be essential in the growth of information networks~\cite{medo2011temporal,medo2014statistical,holme2012temporal,parolo2015attention}. The relationship between the degree of a node and its future degree increase is therefore more complex (see Figure~\ref{fig:dk_vs_k}). Particularly, two nodes of the same degree may have very different expected degree increase if their age differs substantially. In summary, only evaluation on the time probe can be representative of how a recommendation method represents the users' future actions and interests.

With the help of the network growth model with relevance and aging~\cite{medo2011temporal}, the previous arguments can be put in a more precise mathematical form. According to the model, nodes acquire new links at the rate proportional to
\begin{equation}
\label{model}
k_{\alpha}(t) R_{\alpha}(t)
\end{equation}
where $R_{\alpha}(t)$ is the relevance of node $\alpha$ at time $t$; $R_{\alpha}(t)$ generally decays with time. It thus follows that $k_{\alpha}(t)$ alone is a poor predictor of the node degree increase in the future because small relevance can prevent a popular node from receiving many new links. To obtain better predictions of degree increase, one may look at the system at a sufficiently short time scale $\tau$ over which both $k_{\alpha}(t)$ and $R_{\alpha}(t)$ change little and therefore, according to Eq.~(\ref{model}), the node's propensity to attract new links remains approximately unchanged. If the node has attracted $k_{\alpha}(t) - k_{\alpha}(t-\tau)$ links in the past, we thus expect it to attract the same number of links in the future. Recent degree of a node is therefore anticipated to predict well the node's future degree increase.

\section{Recommendation methods and their evaluation}
\label{sec:methods}
We now describe an elementary network-based recommendation method, a benchmark time-aware method, and a novel hybrid method. Two further network-based methods~\cite{zhou2010solving,zeng2014information} are described in SI.

\noindent (1) \emph{Probabilistic spreading} (ProbS). All items collected by a chosen user until time $t$ are assigned a unit amount of resource which is then propagated over the bipartite network by a random walk process from the item side to the user side and back~\cite{zhou2007bipartite}. In each step, the amount of resource at a node is uniformly distributed among the node's neighbors. The resource values obtained on the item side in the second propagation step are consequently interpreted as item recommendation scores for the given user. Denoting the initial resource vector for user $i$ as $\vek{f}^{(i)}$ (its elements are $f_{\alpha}^{(i)} = a_{i\alpha}(t)$), the final resource values $g_{\alpha}^{(i)}$ can be written as
$g_{\alpha}^{(i)} = \sum_{\beta=1}^I W_{\alpha\beta}(t)f_{\beta}^{(i)}$
where
\begin{equation}
\label{ProbSW}
W_{\alpha\beta}(t) = \frac1{k_{\beta}(t)}\sum_{j=1}^U \frac{a_{j\alpha}(t)a_{j\beta}(t)}{k_j(t)}.
\end{equation}
This directly represents the above-described spreading process: $a_{j\alpha}a_{j\beta}$ corresponds to paths going from item $\beta$ to item $\alpha$ through user $j$ and the division with $k_{\beta}$ and $k_j$ corresponds to the uniform division of resources among neighbors of nodes $\beta$ and $j$, respectively.

\noindent (2) \emph{Recent degree increase} (DI) has been shown to perform well in predicting future popularity of items~\cite{zeng2013trend} (see also Figure~\ref{fig:dk_vs_k}d). After choosing the time window length $\tau$, one computes item degree increase within this window as
$\Delta k_{\alpha}(t, \tau) = k_{\alpha}(t) - k_{\alpha}(t - \tau)$. The drawback of this approach is that many items can share the same value of $\Delta k_{\alpha}(t,\tau)$ and the order in which these items should be represented to a user is then ambiguous (this is particularly relevant when $\tau$ is small and many of the degree increase values are small or zero). Since item popularity is an important factor in the evolution of a network, we complement recent degree increase with item popularity and compute item score as
\begin{equation}
\Delta k_{\alpha}'(t, \tau) = \Delta k_{\alpha}(t, \tau) + \varepsilon k_{\alpha}(t).
\end{equation}
The value of $\varepsilon$ is set small enough so that the order of items of different $\Delta k_{\alpha}(t, \tau)$ is not changed (this is guaranteed when $\varepsilon<1/\max_{\alpha} k_{\alpha}$; we use $\varepsilon = 10^{-9}$ in our simulations).

\noindent (3) \emph{Time-aware probabilistic spreading} (TProbS). Items at the top of ProbS's recommendation lists tend to have high degree~\cite{liu2011information,qiu2011item}. To achieve high degree, nodes typically need considerable time and, as a result, ProbS tends to recommend old items. To improve the ranking of recent items, we propose to combine the ProbS score $g_{\alpha}^{(i)}$ with the network's recent temporal dynamics as
\begin{equation}
t_{\alpha}^{(i)} = g_{\alpha}^{(i)}\,\frac{\Delta k_{\alpha}'(t, \tau)}{k_{\alpha}(t)}
\end{equation}
where the division with $k_{\alpha}(t)$ compensates the intrinsic proportionality of ProbS scores to item degree.

\noindent (4) \emph{Similarity-preferential diffusion} (SimS) has been proposed in~\cite{zeng2014information} and found to be one of the best-performing network-based recommendation methods in~\cite{yu2016network}. One first introduces user similarity based on the items that the two users have both collected
\begin{equation}
s_{ij} = \sum_{\beta=1}^I \frac{a_{i\beta}a_{j\beta}}{k_{\beta}}.
\end{equation}
and the final resource values take the form
\begin{equation}
\label{SimS_eq}
h_{\alpha}^{(i)} = \sum_{j=1}^U \frac{a_{j\alpha}s_{ij}^{\theta}}{k_j^{\lambda}k_{\alpha}^{1-\lambda}}
\end{equation}
Here $\theta>0$ is a parameter that decides how much more weight we give the users with high similarity and $\lambda\in[0,1]$ is a parameter that combines ProbS with its counterpart \emph{HeatS} (heat spreading; see~\cite{zhou2010solving} for the ProbS-HeatS hybrid). SimS reduces to ProbS when $\theta=1$ and $\lambda=1$ and to HeatS when $\theta=1$ and $\lambda=0$.

\noindent (5) \emph{Time-aware version of the ProbS-HeatS hybrid method} (THybrid) combines SimS with DI in the same way as TProbS. Denoting the final resource values of the ProbS-HeatS hybrid as $h_{\alpha}^{(i)}$, THybrid score reads
\begin{equation}
u_{\alpha}^{(i)} = h_{\alpha}^{(i)}\,\frac{\Delta k_{\alpha}'(t, \tau)}{k_{\alpha}(t)}
\end{equation}
THybrid has two parameters to optimize over: $\lambda$ and $\tau$.

\begin{figure}
\centering
\includegraphics[scale=0.55]{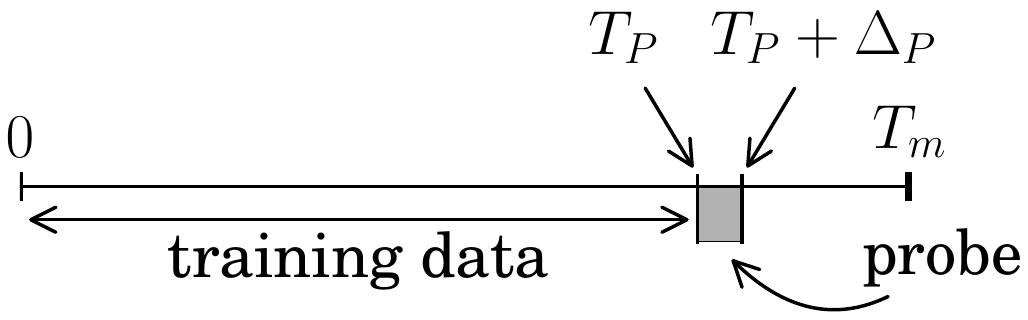}
\caption{Construction of the time probe.}
\label{fig:probe}
\end{figure}

As mentioned above, recommendation methods are evaluated by measuring their ability to reproduce the hidden probe data. The traditional random probe consists of a chosen fraction (we use here 10\%) of all links that are chosen at random. The remaining 90\% of data then form the training data that are used as input for recommendation methods. The natural way to modify this construction to a time-aware probe would be to choose the most recent 10\% of all links. However, this approach has two disadvantages. First, it would only allow to construct one probe (as opposed to the random probe where different random divisions of the input data can be evaluated) and hence provides limited information about the inherent variability of recommendation performance. Second, due to new items constantly appearing in the system, the most recent 10\% of links that are moved to the time-aware probe are likely to contain a number of new items that are not present in the remaining training data. Since they lack any connections, these items cannot be recommended to the users by a network-based recommendation method which effectively and non-transparently reduces the size of the probe set to the actually recommendable items. While various approaches have been developed to overcome this ``cold start'' problem~\cite{schein2002methods}, they generally go beyond the scope of this article.

To avoid the mentioned disadvantages, we choose a different time-aware approach. Assuming that the time stamps of the complete data run from $0$ to $T_m$, we choose the probe time stamp $T_P$ at random and move all entries with time stamps from $[T_P, T_P + \Delta_P)$ to the probe. All entries with time stamps smaller than $T_P$ then constitute the training data (see Figure~\ref{fig:probe} for an illustration). The parameter $\Delta_P$ characterizes the temporal span of the probe; it should be long enough to allow a substantial number of entries to enter the probe, yet short enough to prevent a large number of probe entries correspond to items that are not present in the training data. We choose $\Delta_P$ one day for Netflix and Yelp, and one hour for Digg. To calibrate the recommendation methods, we constraint the choice of $T_P$ to $[0.8T_m, 0.9T_m]$. To obtain the final performance measurements, we draw $T_P$ at random from $[0.9T_m, T_m - \Delta_P]$. We use $\Delta_P=1\,\text{day}$ for Netflix and Yelp and $\Delta_P=1\,\text{hour}$ for Digg. The exact choice of $\Delta_P$ and the $T_P$ ranges does not qualitatively alter the results. In both steps, results are averaged over 100 values of $T_P$.

\subsection{Evaluation metrics}
Upon computing recommendation scores of items for all users, we are ultimately interested in knowing, which items we should ``recommend'' to each individual user. Since all methods implicitly assume that the higher the score, the more likely the item is to be appreciated by the user, a recommendation list for a user is obtained by ranking all items according to their recommendation score for the given user in a descending order. In doing so, the items that have been already collected by the user are ignored because the user already knows them.

To quantify recommendation accuracy, we use recall which is a standard metric for accuracy in information filtering~\cite{herlocker2004evaluating,shani2011evaluating}. For a given user, recall is defined as the fraction of the user's probe items that occur in top $L$ places of the recommendation list. Overall recall $R(L)$ is then computed by averaging recall values over all users. Recall ranges from zero (worst) to one (best). The ranking score of item $\alpha$ in the recommendation list of user $i$ is obtained by dividing the item's ranking position $r_{i\alpha}$ with the number of items that have not yet been collected by user $i$. Ranking score of a method is obtained by averaging $r_{i\alpha} / (I - k_i)$ over all probe entries~\cite{zhou2010solving}. The best result is close to zero (when all probe items rank at the top of their respective recommendation lists, indicating good recommendation results), while the worst result is 0.5 (when the probe items are randomly distributed in the recommendation lists). Finally, to address the diversity aspect of recommendation~\cite{zhou2010solving}, we compute the average degree (popularity) of the top $L$ items at time $T_P$, $k_R(L)$. We use $L=50$ for $R(L)$ and $k_R(L)$.

\begin{table}
\centering\setlength{\tabcolsep}{6pt}
\begin{tabular}{lrrrr}
\hline
\hline
Dataset &     $U$ &    $I$ &       $E$ & time span\\
\hline
Netflix &   1,999 & 12,111 &   352,394 & 2,187 days\\
Yelp    & 123,368 & 41,958 &   674,758 & 3,558 days\\
Digg    &  68,480 &  3,553 & 2,337,418 & 35 days\\
\hline
\hline
\end{tabular}
\caption{Basic properties of the used datasets: the number of users $U$, the number of items $I$, the number of links $E$, and the total time span.}
\label{tab:datasets}
\end{table}

\subsection{Description of the analyzed datasets}
We evaluate the chosen recommendation methods on three datasets originating from distinct e-commerce systems. Basic statistical properties of these datasets are summarized in Table~\ref{tab:datasets}.

\url{Netflix.com} is a streaming service that was mainly a DVD rental service at the time when the dataset has been released. The company has released the dataset, which is nowadays a golden standard in the information filtering community, for its Netflix Prize~\cite{bennett2007netflix}. Out of the 450,000 users contained in the original dataset, we have chosen 2000 users at random and connect them with the items that they have rated with rating 3 or higher. Resolution of the dataset's time stamps is one day.

\url{Yelp.com} is a website where users can review and rate various businesses such as restaurants, doctors, and bars. This and several other datasets have been released by the company for its dataset challenges~\cite{yelp_dataset}. In the dataset, the users are connected with the businesses that they have rated with rating 3 or higher. Resolution of the dataset's time stamps is one day.

In \url{Digg.com}, users can vote for any story of their interest. The Digg dataset is composed of news that were collected over the period of one month in 2009~\cite{lerman2012social}; only stories that were promoted to the web site's front page are included. While the mechanism of promotion to the font page is not explicitly available, it is obvious that story popularity is an important factor (the lowest story degree in the dataset is 111). A link is created in the network representation between a user and a story when the user votes for the story. Resolution of the dataset's time stamps is one minute.

\begin{figure*}
\centering
\includegraphics[scale=0.63]{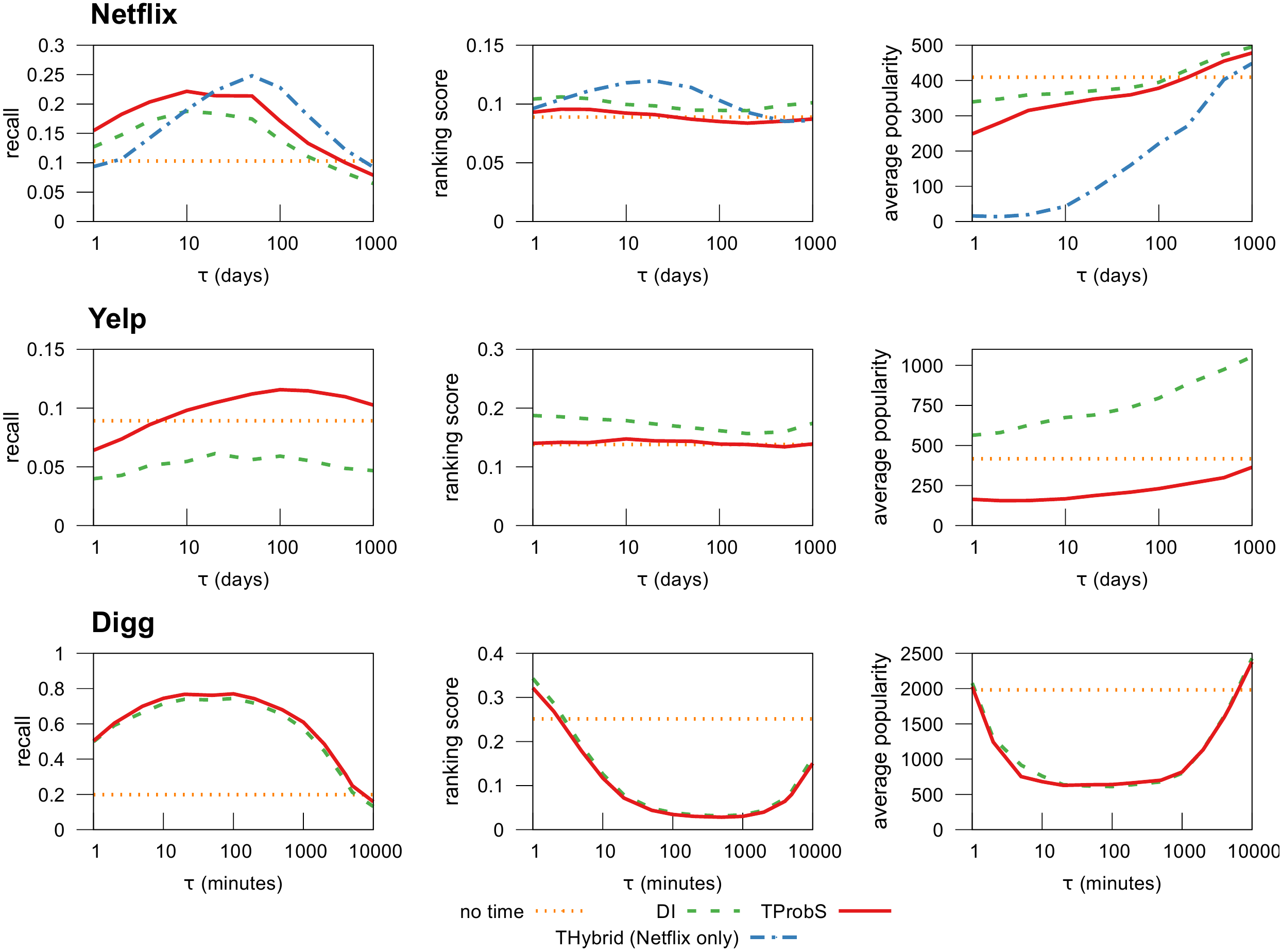}
\caption{Method calibration on time probes with $T_P\in[0.8T_m, 0.9T_m]$: Recommendation recall, ranking score, and popularity of the recommended items as a function of the window length $\tau$ for the evaluated methods. Horizontal lines mark the performance of the time-unaware method with the highest recall (SimS with $\lambda=0.3$ and $\theta=1.25$ for Netflix, ProbS for Yelp, and HeatS for Digg).}
\label{fig:tau_dependence}
\end{figure*}

\section{Results}
Method calibration with respect to the time window length $\tau$ is shown in Figure~\ref{fig:tau_dependence} (for methods with multiple parameters, we show here their best-performing choice). The recall-maximizing parameter values are then used in out-of-sample measurement ($T_P\in[0.9T_m, T_m - \Delta_P]$) of recommendation performance that is summarized in Table~\ref{tab:results}. This table further includes recall values obtained with ProbS on the random probe. In agreement with our previous arguments about the inadequacy of the random probe for recommendation evaluation, we see that the recall values drop by the factor two or three once the time probe is used instead of the random one. The random probe gives an excessively optimistic view of the performance of network-based recommendation methods.

\begin{table*}
\centering
\begin{tabular}{rccccc}
\hline
\hline
& random probe & \multicolumn{4}{c}{time probe}\\
\cline{3-6}
Dataset & ProbS & ProbS &   DI & No time & TProbS\\
\hline
Netflix &         0.24 &  0.08 & 0.15 & 0.10 & {\bfseries 0.18}\\
Yelp    &         0.18 &  0.09 & 0.06 & 0.10 & {\bfseries 0.12}\\
Digg    &         0.18 &  0.06 & 0.71 & 0.15 & {\bfseries 0.74}\\
\hline
\hline
\end{tabular}
\caption{Method evaluation on time probes with $T_P\in[0.9T_m, T_m - \Delta_P]$: Recall for recommendation list length $L=50$ for ProbS on the random probe and for all evaluated methods on the time probe; we use here the optimal parameter values presented in Figure~\ref{fig:tau_dependence}. ``No time'' stands for the best-performing time-unaware recommendation method (specified in the caption of Figure~\ref{fig:tau_dependence}). For the time probe, TProbS is the best among the shown methods in all three datasets. For Netflix, THybrid further improves recall to $0.19$.}
\label{tab:results}
\end{table*}

The main finding is that the use of time-aware recommendation substantially improves recall in all three datasets studied here. The biggest recall improvement is observed on the Digg data where aging is much faster than in Netflix and Digg (popular Digg stories reach 50\% of their final popularity on average in 16 hours as opposed to 681 days for Netflix and 691 days for Yelp). At the same time, recall obtained in Digg data with time-aware recommendation is only marginally higher than recall obtained with recent degree increase. This is due to the selective nature of this dataset which includes only the most popular Digg stories. Heterogeneous taste of users and personalized recommendation are thus limited in effect. We present the Digg dataset because of its quick aging dynamics that fully exposes the inappropriateness of traditional recommendation methods. By contrast, the gap between TProbS and DI is particularly big for the Yelp data, which may be due to geographical effects: even when a restaurant's recent degree increase is high, recommending it to a user who lives far away is unlikely to prove useful. By contrast, TProbS does not face the same problem because user location is reflected by their collection patterns. One can also note that the relative improvement achieved with TProbS on this dataset is smaller than it is on the Netflix and Digg data. An inspection of the time probe reveals that this is due to the fact that in the Yelp data, the aging dynamics is somewhat different as there are items of substantial age that are present in the probe often (perpetually popular items).

Besides significantly improving recommendation recall for all three datasets, the average degree of the recommended items is for TProbS considerably lower than for time-unaware methods which demonstrates the method's potential to increase recommendation diversity which is a long-standing issue~\cite{ziegler2005improving,zhang2008avoiding,adomavicius2012improving}. In particular, when TProbS is used instead of ProbS, $k_R(50)$ decreases by 38\%, 50\%, and 82\% for the Netflix, Yelp, and Digg data, respectively. THybrid, which is a time-aware modification of the popular ProbS-HeatS hybrid recommendation method, further improves recall and lowers the average popularity of top-ranked items. Results for other recommendation list lengths $L$ are qualitatively the same.

Figure~\ref{fig:tau_dependence} further shows that for the Netflix and Yelp data, the ranking score achieved by TProbS is nearly independent of $\tau$, which is very different from recall that strongly changes with $\tau$. This is an indication that the ranking score with its coverage of all probe items---including those that are ranked very badly in users' recommendation lists---is a little reliable indicator of recommendation performance. By contrast, recall is only sensitive to probe items appearing at the top $L$ places of recommendation lists and thus has the potential to reflect the users' benefit from recommendation.

\section{Discussion}
\label{sec:discussion}
By contrast to other recommendation approaches~\cite{ding2005time,koren2010collaborative,campos2014time,daneshmand2014time}, network-based recommendation methods do not pay attention to time stamps in the input data and are commonly evaluated on the random probe. We demonstrate that the evaluation on the random probe is fundamentally flawed as: (1) it essentially measures how a method reproduces the network topology of the input data and (2) tends to overestimate the method's performance in realistic settings. By contrast, the evaluation based on the time probe measures the method's ability to reflect both network topology as well as the system's natural growth patterns and the users' shifting interests.

Facing the lowered recommendation performance of network-based methods under the new evaluation protocol, we propose to enhance the existing time-unaware methods with temporal features and show that even their simple combination dramatically improves the resulting recommendation accuracy. The improvement is also with respect to using temporal features only which suggests that to obtain the best performance, it is crucial to combine temporal features and personalized preferences. Besides improved accuracy, we find that time-aware methods tend to recommend items of smaller degree than traditional methods and therefore can be used to improve the degree of novelty and serendipity (the ability to produce unexpected, yet useful results) in recommendation~\cite{herlocker2004evaluating,ge2010beyond}. Our work opens the door to the search for even more effective time-aware methods for recommendation and information filtering in general.

\acknowledgments
This work was supported by the Swiss National Science Foundation Grant No. 200020-143272 and by the EU FET-Open Grant No. 611272 (project Growthcom).


\begin{thebibliography}{10}
\expandafter\ifx\csname url\endcsname\relax\def\url#1{\texttt{#1}}\fi

\bibitem{lazer2009life}
\Name{Lazer D. \etal} \REVIEW{Science}{323}{2009}{721}.

\bibitem{dong2012link}
\Name{Dong Y. \etal} \Book{Link prediction and recommendation across
  heterogeneous social networks} in proc. of \Book{12th IEEE International
  Conference on Data Mining} (IEEE) 2012 pp. 181--190.

\bibitem{bao2015recommendations}
\Name{Bao J., Zheng Y., Wilkie D. \and Mokbel M.}
  \REVIEW{GeoInformatica}{19}{2015}{525}.

\bibitem{schafer1999recommender}
\Name{Schafer J.~B., Konstan J. \and Riedl J.} \Book{Recommender systems in
  e-commerce} in proc. of \Book{1st ACM conference on Electronic commerce}
  (ACM) 1999 pp. 158--166.

\bibitem{bell2007lessons}
\Name{Bell R.~M. \and Koren Y.} \REVIEW{ACM SIGKDD Explorations
  Newsletter}{9}{2007}{75}.

\bibitem{pathak2010empirical}
\Name{Pathak B., Garfinkel R., Gopal R.~D., Venkatesan R. \and Yin F.}
  \REVIEW{Journal of Management Information Systems}{27}{2010}{159}.

\bibitem{bobadilla2013recommender}
\Name{Bobadilla J., Ortega F., Hernando A. \and Guti{\'e}rrez A.}
  \REVIEW{Knowledge-Based Systems}{46}{2013}{109}.

\bibitem{newman2010networks}
\Name{Newman M.} \Book{Networks: An introduction} (Oxford University Press)
  2010.

\bibitem{zhang2007recommendation}
\Name{Zhang Y.-C. \etal} \REVIEW{EPL}{80}{2007}{68003}.

\bibitem{zhou2007bipartite}
\Name{Zhou T., Ren J., Medo M. \and Zhang Y.-C.} \REVIEW{Physical Review
  E}{76}{2007}{046115}.

\bibitem{zhou2010solving}
\Name{Zhou T. \etal} \REVIEW{Proceedings of the National Academy of
  Sciences}{107}{2010}{4511}.

\bibitem{yu2016network}
\Name{Yu F., Zeng A., Gillard S. \and Medo M.} \REVIEW{Physica
  A}{452}{2016}{192}.

\bibitem{ding2005time}
\Name{Ding Y. \and Li X.} \Book{Time weight collaborative filtering} in proc.
  of \Book{Proceedings of the 14th ACM international conference on Information
  and knowledge management} (ACM) 2005 pp. 485--492.

\bibitem{koren2010collaborative}
\Name{Koren Y.} \REVIEW{Communications of the ACM}{53}{2010}{89}.

\bibitem{campos2014time}
\Name{Campos P.~G., D{\'\i}ez F. \and Cantador I.} \REVIEW{User Modeling and
  User-Adapted Interaction}{24}{2014}{67}.

\bibitem{daneshmand2014time}
\Name{Daneshmand S.~M., Javari A., Abtahi S.~E. \and Jalili M.} \REVIEW{The
  Computer Journal}{58}{2015}{1955}.

\bibitem{crane2008robust}
\Name{Crane R. \and Sornette D.} \REVIEW{Proceedings of the National Academy of
  Sciences}{105}{2008}{15649}.

\bibitem{szabo2010predicting}
\Name{Szabo G. \and Huberman B.~A.} \REVIEW{Communications of the
  ACM}{53}{2010}{80}.

\bibitem{ren2016characterizing}
\Name{Ren Z.-M., Shi Y.-Q. \and Liao H.} \REVIEW{Physica A: Statistical
  Mechanics and its Applications}{453}{2016}{236}.

\bibitem{medo2011temporal}
\Name{Medo M., Cimini G. \and Gualdi S.} \REVIEW{Physical Review
  Letters}{107}{2011}{238701}.

\bibitem{medo2014statistical}
\Name{Medo M.} \REVIEW{Physical Review E}{89}{2014}{032801}.

\bibitem{hou2014memory}
\Name{Hou L., Pan X., Guo Q. \and Liu J.-G.} \REVIEW{Scientific
  reports}{4}{2014}{}.

\bibitem{wang2013quantifying}
\Name{Wang D., Song C. \and Barab{\'a}si A.-L.}
  \REVIEW{Science}{342}{2013}{127}.

\bibitem{zhou2015temporal}
\Name{Zhou Y., Zeng A. \and Wang W.-H.} \REVIEW{PLoS ONE}{10}{2015}{}.

\bibitem{mariani2016quantifying}
\Name{Mariani M., Medo M. \and Zhang Y.-C.} \Book{Quantifying the significance
  of scientific papers through time-balanced network centrality} (under
  review).

\bibitem{herlocker2004evaluating}
\Name{Herlocker J.~L., Konstan J.~A., Terveen L.~G. \and Riedl J.~T.}
  \REVIEW{ACM Transactions on Information Systems}{22}{2004}{5}.

\bibitem{shani2011evaluating}
\Name{Shani G. \and Gunawardana A.} \Book{Evaluating recommendation systems} in
  \Book{Recommender systems handbook} (Springer) 2011 pp. 257--297.

\bibitem{lu2012recommender}
\Name{L{\"u} L., Medo M., Yeung C.~H., Zhang Y.-C., Zhang Z.-K. \and Zhou T.}
  \REVIEW{Physics Reports}{519}{2012}{1}.

\bibitem{zeng2013trend}
\Name{Zeng A., Gualdi S., Medo M. \and Zhang Y.-C.} \REVIEW{Advances in Complex
  Systems}{16}{2013}{1350024}.

\bibitem{clauset2009power}
\Name{Clauset A., Shalizi C.~R. \and Newman M.~E.} \REVIEW{SIAM
  Review}{51}{2009}{661}.

\bibitem{zhou2011emergence}
\Name{Zhou T., Medo M., Cimini G., Zhang Z.-K. \and Zhang Y.-C.} \REVIEW{PLoS
  One}{6}{2011}{e20648}.

\bibitem{liu2011information}
\Name{Liu J.-G., Zhou T. \and Guo Q.} \REVIEW{Physical Review
  E}{84}{2011}{037101}.

\bibitem{qiu2011item}
\Name{Qiu T., Chen G., Zhang Z.-K. \and Zhou T.} \REVIEW{EPL}{95}{2011}{58003}.

\bibitem{albert2002statistical}
\Name{Albert R. \and Barab{\'a}si A.-L.} \REVIEW{Reviews of Modern
  Physics}{74}{2002}{47}.

\bibitem{holme2012temporal}
\Name{Holme P. \and Saram{\"a}ki J.} \REVIEW{Physics Reports}{519}{2012}{97}.

\bibitem{parolo2015attention}
\Name{Parolo P. D.~B., Pan R.~K., Ghosh R., Huberman B.~A., Kaski K. \and
  Fortunato S.} \REVIEW{Journal of Informetrics}{9}{2015}{734}.

\bibitem{zeng2014information}
\Name{Zeng A., Vidmer A., Medo M. \and Zhang Y.-C.}
  \REVIEW{EPL}{105}{2014}{58002}.

\bibitem{schein2002methods}
\Name{Schein A.~I., Popescul A., Ungar L.~H. \and Pennock D.~M.} \Book{Methods
  and metrics for cold-start recommendations} in proc. of \Book{Proceedings of
  the 25th annual international ACM SIGIR conference on Research and
  development in information retrieval} (ACM) 2002 pp. 253--260.

\bibitem{bennett2007netflix}
\Name{Bennett J. \and Lanning S.} \Book{The netflix prize} in proc. of
  \Book{Proceedings of KDD cup and workshop} Vol. 2007 2007 p.~35.

\bibitem{yelp_dataset}
\Name{{Yelp Inc.}} \Book{Yelp dataset challenge} obtained from
  https://www.yelp.com/dataset\_challenge (2014).

\bibitem{lerman2012social}
\Name{Lerman K., Ghosh R. \and Surachawala T.} \REVIEW{arXiv preprint
  arXiv:1202.3162}{}{2012}{}.

\bibitem{ziegler2005improving}
\Name{Ziegler C.-N., McNee S.~M., Konstan J.~A. \and Lausen G.} \Book{Improving
  recommendation lists through topic diversification} in proc. of
  \Book{Proceedings of the 14th international conference on World Wide Web}
  (ACM) 2005 pp. 22--32.

\bibitem{zhang2008avoiding}
\Name{Zhang M. \and Hurley N.} \Book{Avoiding monotony: improving the diversity
  of recommendation lists} in proc. of \Book{Proceedings of the 2008 ACM
  conference on Recommender systems} (ACM) 2008 pp. 123--130.

\bibitem{adomavicius2012improving}
\Name{Adomavicius G. \and Kwon Y.} \REVIEW{IEEE Transactions on Knowledge and
  Data Engineering}{24}{2012}{896}.

\bibitem{ge2010beyond}
\Name{Ge M., Delgado-Battenfeld C. \and Jannach D.} \Book{Beyond accuracy:
  evaluating recommender systems by coverage and serendipity} in proc. of
  \Book{Proceedings of the fourth ACM conference on Recommender systems} (ACM)
  2010 pp. 257--260.
\end{thebibliography}
\end{document}